\PassOptionsToClass{10pt}{revtex4-2}
\documentclass[aps,prl,showpacs,floatfix,twocolumn,byrevtex,superscriptaddress]{revtex4-1}

\usepackage[utf8]{inputenc}
\usepackage{amsmath}
\usepackage{amssymb}
\usepackage{amstext}
\usepackage{amsopn}
\usepackage{amsfonts}
\usepackage{amsxtra}
\usepackage{fixltx2e}
\usepackage[english]{babel}
\usepackage{graphicx}
\usepackage{float}
\usepackage{bm}
\usepackage{multirow}
\usepackage{dcolumn}
\usepackage{color}
\usepackage{hyperref}
\usepackage{todonotes}
\usepackage{verbatim}
\usepackage{soul}
\usepackage{xcolor}

\def\MBT{MnBi$_2$Te$_4$}

\def\MBST{MnBi$_{2-x}$Sb$_x$Te$_4$}
\def\MBSTpsev{MnBi$_{1.3}$Sb$_{0.7}$Te$_4$}

\date{\today}

\begin{document}
\title{On the Role of Defects in the Electronic Structure of \MBST{}}
\author{D. Nevola}\email[]{nevola@bnl.gov}
\affiliation{Condensed Matter Physics and Materials Science Department, Brookhaven National Laboratory, Upton, New York 11973, USA}
\author{Kevin F. Garrity}\email[]{kevin.garrity@nist.gov}
\affiliation{Materials Measurement Laboratory, National Institute for Standards and Technology, Gaithersburg, MD 20899, USA}
\author{N. Zaki}
\affiliation{Condensed Matter Physics and Materials Science Department, Brookhaven National Laboratory, Upton, New York 11973, USA}
\author{J.-Q. Yan}
\affiliation{Materials Science and Technology Division, Oak Ridge National Laboratory, Oak Ridge, Tennessee 37831, USA}
\author{H. Miao}\email[]{miaoh@ornl.gov}
\affiliation{Materials Science and Technology Division, Oak Ridge National Laboratory, Oak Ridge, Tennessee 37831, USA}
\author{Sugata Chowdhury}\email[]{sugata.chowdhury@howard.edu}
\affiliation{Materials Measurement Laboratory, National Institute for Standards and Technology, Gaithersburg, MD 20899, USA; Department of Physics and Astronomy, Howard University, 2355 6th St NW, Washington, DC 20059, USA}
\author{P. D. Johnson}\email[]{pdj@bnl.gov}
\affiliation{Condensed Matter Physics and Materials Science Department, Brookhaven National Laboratory, Upton, New York 11973, USA}

\begin{abstract}

Elemental substitution is a proven method of Fermi level tuning in topological insulators, which is needed for device applications. Through static and time resolved photoemission, we show that in \MBT{}, elemental substitution of Bi with Sb indeed tunes the Fermi level towards the bulk band gap, making the material charge neutral at 35\% Sb concentration. For the first time, we are able to directly probe the excited state band structure at this doping level, and their dynamics, which show that the decay channels at the Fermi level are severely restricted. However, elemental substitution widens the surface state gap, which we attribute to the increase in antisite defects resulting from Sb substitution. This hypothesis is supported by DFT calculations that include defects, which show a sensitivity of the topological surface state to their inclusion. Our results emphasize the need for defect control if \MBSTpsev{} is to be used for device applications.
\end{abstract}

\maketitle


The interplay between magnetism and topology in magnetic topological insulator materials offer exciting opportunities to realize exotic phenomena such as Majorana modes\cite{Yang2019}, the axion insulator phase\cite{Yufei2021,liu2020}, or the quantum anomolous Hall effect (QAHE)\cite{liu2020,Deng2020,deng2021,Mong2010,He1093}. \MBT{} (MBT) has been identified as the first intrinsic magnetic topological insulator (MTI), exhibiting A-type antiferromagnetic (AFM) order and a N\'eel temperature of 25~K\cite{Gong2019,Otrokov2019}. Soon after its experimental discovery, the QAHE was demonstrated on few layer MBT\cite{Zhang2020,Deng2020}. However, several challenges emerged that would complicate the realization of the QAHE in the bulk. First, angle resolved photoemission spectroscopy (ARPES) revealed that it possessed heavy intrinsic n-doping\cite{Otrokov2019,Li2019,Chen2019,Hao2019,Lee2019,SwatekPRB2020,Nevola2020,ma2020,Shikin2020,Shikin2021,Garnica2022}. Applying a high gate voltage is then necessary to move the chemical potential inside the bulk band gap, posing challenges for device applications. Secondly, there have been conflicting ARPES reports on whether the topological surface state (TSS) is gapped at the Dirac point or gapless in the paramagnetic phase\cite{Otrokov2019,Li2019,Chen2019,Hao2019,Lee2019,SwatekPRB2020,Nevola2020,ma2020,Shikin2020,Shikin2021,Garnica2022}. Although various explanations have been proposed\cite{Chen2019,Lee2019,Garrity2021}, the recent observation that there appears to be a sample dependent gap points towards the role of defects\cite{Shikin2021,Garnica2022}. Finally, there are conflicting reports surrounding the opening of the surface gap with the emergence of magnetic order: many studies showing a minimal to zero change in the gap below T$_N$\cite{Garnica2022,SwatekPRB2020,Nevola2020,Otrokov2019,Chen2019,Hao2019,Ji2021,hu2021}. However, the intensity of the TSS increases below T$_N$, suggesting that the magnetic order may effect the penetration depth of the surface state into the bulk\cite{Nevola2020,Estyunin2020}. 

Recently, the focus on MBT has been on determining the role of defects, as it potentially provides an explanation for the unresolved issues discussed above. Several studies have identified that the Bi$_{Mn}^+$ or Mn$_{Bi}^-$ antisite defects (which would indicate excess Bi and Mn respectively) are by far the most abundant, present on the order of a few percent \cite{du2021,Yuan2020,Hou2020,Huang2020,Garnica2022,lai2021,Liu2021defect}. It is thought that they play a role in both the intrinsic n-doping and penetration depth of the topological surface state\cite{Garnica2022,Shikin2021}. Additionally, these defects are also magnetic, which in itself has attracted considerable attention. In particular, it is now known that MBT is a ferrimagnet with the defect layers aligning antiferromagnetically with the original Mn layer\cite{Liu2021defect,lai2021,hu2021tuning,riberolles2021evolution}.

\begin{figure}[tb]
\includegraphics[width=8.6 cm]{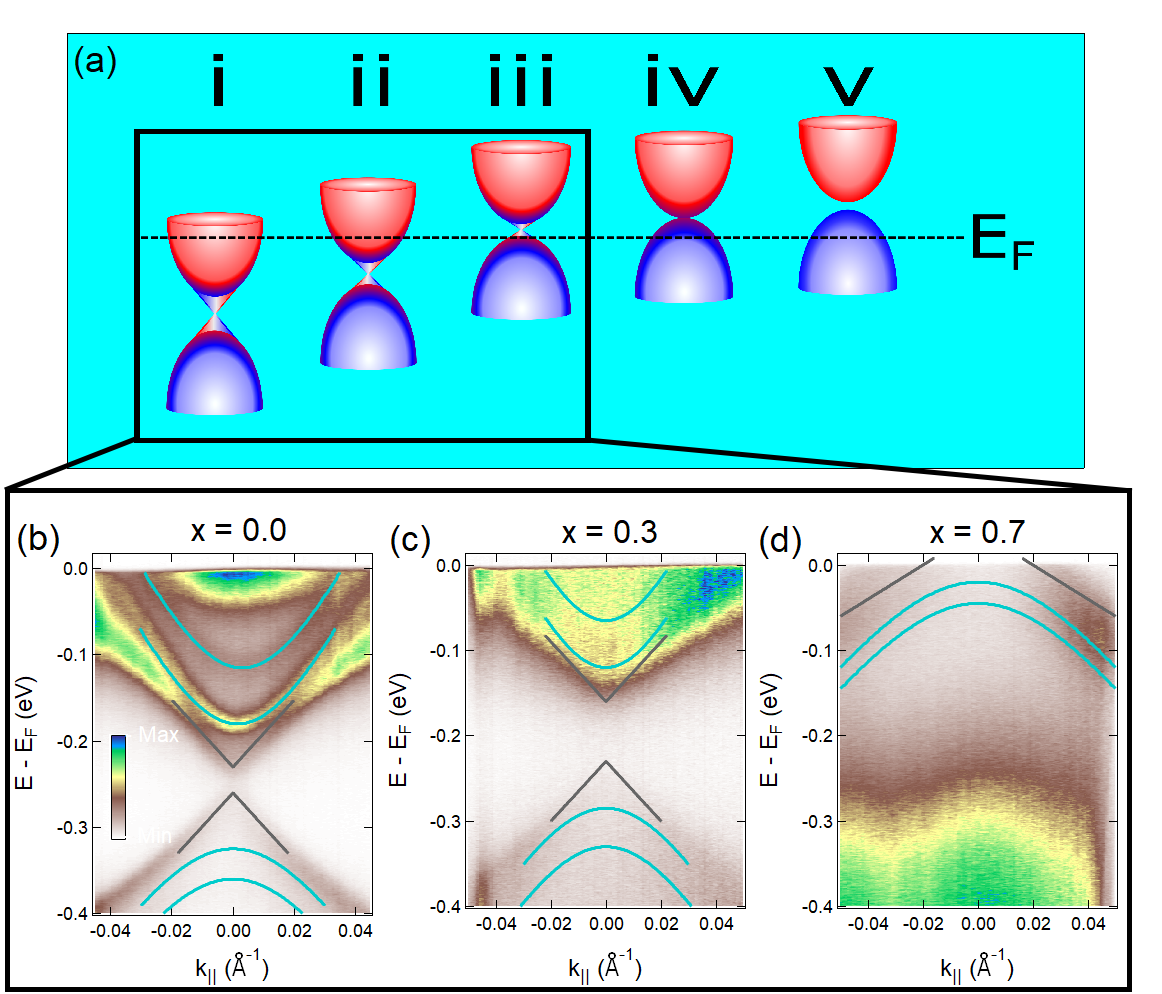}
\caption{(a) Schematic of the expected \MBST{} band structure as a function of Sb doping level, where (i) and (v) represent the two limits of a topological and trivial insulator, respectively. The red and blue states represent the bulk bands and the cones in between them represent the topological surface states. (b)-(d) show the measured high-resolution ARPES structure at 5~K that were schematically represented in i-iii. The blue and gray lines are guides for the reader that show the bulk bands and topological surface states, respectively. For more information on the band dispersions, see Supplemental materials.}
\label{Fig1}
\end{figure}

Chemical substitution of Bi with Sb has proved useful in tuning the Fermi level of nonmagnetic topological insulators such as Bi$_2$Se$_3$ or Bi$_2$Te$_3$ without significantly altering the topological properties\cite{Kong2011,Neupane2012,Arakane2012}. Recent experimental work has shown a similar effect in \MBST{} (MBST), with a bulk insulating material near $\sim30\%$ Sb concentration\cite{BoChen2019,Ma2021,Yan2019,Ko2020,riberolles2021evolution,Ryota2022}. However, there are two complications. First there are conflicting reports on whether the reduced spin-orbit coupling in Sb leads to a topological phase transition (TPT), even though those studies that predict a phase transition do so at a concentration higher than that of the charge neutral material\cite{Wimmer2021,Eremeev2021,Ma2021,Ko2020}. The latter case is schematically depicted in figure \ref{Fig1}a, where the two effects of p-doping and a TPT occur simultaneously. Second, recent experimental work demonstrates an increased surface gap with increased Sb concentration, posing a problem for potential applications\cite{Ma2021,Ko2020}. This may be related to the increased antisite defect concentration resulting from the similar size of Sb with Mn, which reduces their formation energies\cite{du2021,riberolles2021evolution}. Thus, it is important to study the effects of antisite defects on the band structure, and for their control, so that MBST can successfully be utilized in QAHE applications.

Here, we use high resolution, static ARPES and time resolved ARPES (trARPES) to directly show that at a 35\% Sb concentration, the Fermi level lies inside the bulk band gap, making it the ideal doping level for device applications. While our static ARPES directly images the bands below E$_F$, our trARPES images the conduction band $\sim200$~meV above E$_F$ and demonstrates a significantly longer population lifetime than in MBT. However, we also observe significant band broadening and an enhanced topological gap that are likely due to the enhanced antisite defects that accompanies Sb doping. Investigating these experimental observations, we perform DFT calculations that include antisite defects, and find that the bulk band gap and topological surface states vary greatly depending on their type and location.

Both our laser-based static and time-resolved photoemission spectra were obtained using the same SES-2002 electron analyzer. All samples were cleaved in-situ and the base pressure was $1~$x$~10^{-11}$~torr. For details on the optical configurations, see references \cite{Zaki2021} and \cite{Nevola2020}. The Fermi level was carefully calibrated against polycrystalline Au prior to each experiment and the intensity adjusted to minimize space charge. The energy resolution for the static ARPES was 4~meV. The energy and time resolutions for the trARPES ranged from $40-80$~meV and $100-200$~fs depending on the exact optical configurations and pass energies used. 

Single crystals of \MBST{} were grown out of a Bi(Sb)-Te flux. For further details on the sample growth and characterization, see Ref.\cite{Yan2019}.

To study bulk and defect electronic structures, we perform first principles density functional theory calculations\cite{hohenberg1964inhomogeneous,kohn1965self}. We perform surface calculations with defects using truncated tight-binding Wannier Hamiltonians\cite{pizzi2020wannier90} from bulk defect calculations. Further computational details are specified in the supporting information.

\begin{figure*}[tb]
\includegraphics[width=17.2 cm]{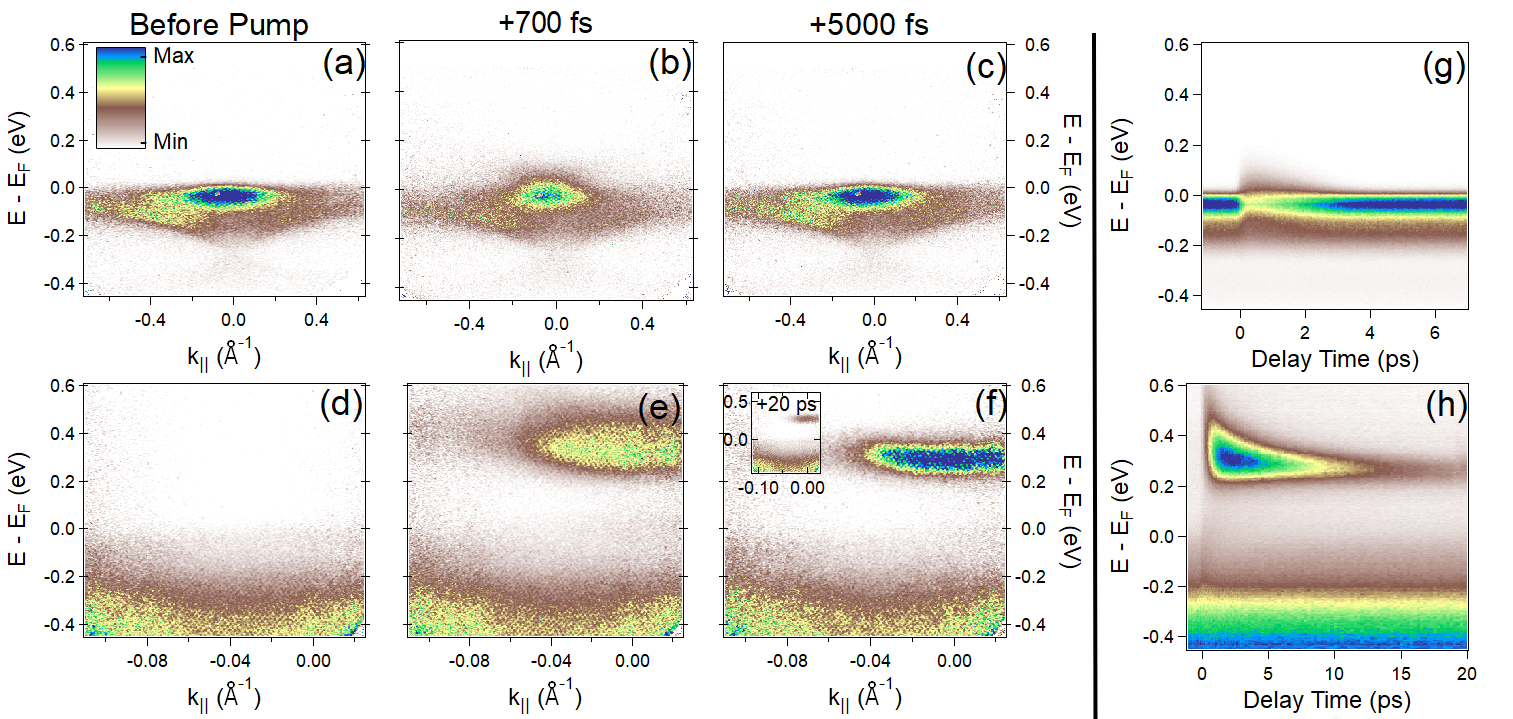}
\caption{trARPES results on the undoped (top row) and $x=0.7$ (bottom row) compounds taken at 30~K. (a)-(c) and (e)-(g) show the band structure at various delay times for the two doping levels. The inset in (g) shows the decay at $\Delta$t~$=+20$~ps. (d) and (h) show the corresponding k-integrated decays.}
\label{Fig2}
\end{figure*}

Our high-energy resolution ARPES results at 5~K are shown in Figs.~\ref{Fig1}b-d. Fig.~\ref{Fig1}b shows the result for the MBT ($x=0$), and are similar with those previously reported\cite{Hao2019,SwatekPRB2020,Chen2019,Shikin2021,Shikin2020}. We observe an intrinsic n-doping, with E{$_F$} located a few hundred meV above the Dirac point (DP) and a 30~meV surface gap. We also observe a bulk conduction band and bulk valence band splitting below T{$_N$}, consistent with previous high resolution studies\cite{Chen2019,SwatekPRB2020,Shikin2020,Shikin2021}. However, the bulk conduction band splitting observed here is 65~meV, slightly higher than that previously reported\cite{Chen2019,Shikin2021}.

Figs.\ref{Fig1}c,d show the ARPES spectra at doping levels of $x=0.3$ and $x=0.7$ respectively. We notice the expected chemical potential shift towards the bulk gap with an increase in Sb concentration. At a doping level of $x=0.3$, the shift is $\sim50$~meV, so that E$_F$ lies just below the Rashba-like state\cite{Nevola2020}, while at a doping level of $x=0.7$, the shift is $\sim250$~meV with respect to the undoped compound, placing E$_F$ just below the DP. As discussed below, we focus our pump-probe experiment on this doping level as E$_F$ lies inside the bulk band gap.

Before we move on to discuss the trARPES results, we note some more subtle points about the spectra presented in Fig.~\ref{Fig1}. First, there is a $\sim70~meV$ surface gap in the doped material, larger than in the undoped material (Fig.~\ref{Fig1}c). (Fig.~\ref{Fig1}b). This agrees with previously reported $\mu$-ARPES results on these doped compounds, where a gap enhancement in the surface state was observed with doping\cite{ma2020} and attributed to impurity scattering. Second, there is a broadening of the bands with doping, consistent with disorder. Although we are unable to clearly resolve the three bands due to the broadening in figure~\ref{Fig1}c (two originating from the spin-split bulk bands and the third due to the surface state), we are able to resolve differences between the 30~K and 5~K spectra by performing additional analysis, including taking energy cuts and second derivatives, signifying that magnetic order has a similar effect on the bulk band structure that it has on undoped MBT[Supplemental].

The band broadening creates an additional difficulty in the $x=0.7$ doping level because of the difficulty involved in determining whether E$_F$ lies inside the bulk gap, or the bulk valence band from the static ARPES alone. However, we do observe a broad peak at $k=0$ centered $\sim$30~meV below E$_F$[Supplemental], suggesting that it may be bulk insulating. Thus, we turn to trARPES, where this information can be revealed by observing the relaxation channels. For example, in the nonmagnetic TI's, the relaxation times drastically increase as E$_F$ is tuned inside the gap because of the relaxation bottleneck\cite{Sumida2017,Zhu2015}. 


The dynamics at two different doping levels is shown in figure~\ref{Fig2}. In order to avoid spectral weight changes due to magnetic order\cite{Nevola2020}, we report results taken above T$_N$. Figures~\ref{Fig2}a-c and \ref{Fig2}d-f show the band structures of MBT and \MBSTpsev{} respectively, at three delay times after the initial pump. We make three important observations. First, we observe the bottom of the conduction band situated at 200~meV above the Fermi level. This value is consistent with both scanning tunneling spectroscopy measurements and calculations\cite{Ko2020,Ma2021}. However, the conduction band is again significantly broadened with respect to the undoped compound (Fig.~\ref{Fig1}b), likely caused by a combination of disorder and the reduced energy resolution in the pump probe study. Secondly, we do not clearly observe a Dirac cone in the doped sample. This is consistent with our static ARPES measurement (Figure~\ref{Fig1}c), where Sb doping gaps the TSS. It is likely that the TSS is gapped here, but our reduced energy resolution makes it difficult to observe. Thirdly, the decay timescale for \MBSTpsev{} is nearly an order of magnitude larger than for MBT, as shown in figures \ref{Fig2}g,h. In the undoped sample, the particles return back to their equilibrium states after $\sim4$~ps, while in the $x=0.7$ case, there is a prominent occupation at the highest measured delay time of 20~ps (Fig.\ref{Fig2}f inset).

\begin{figure}[tb]
\includegraphics[width=8.6 cm]{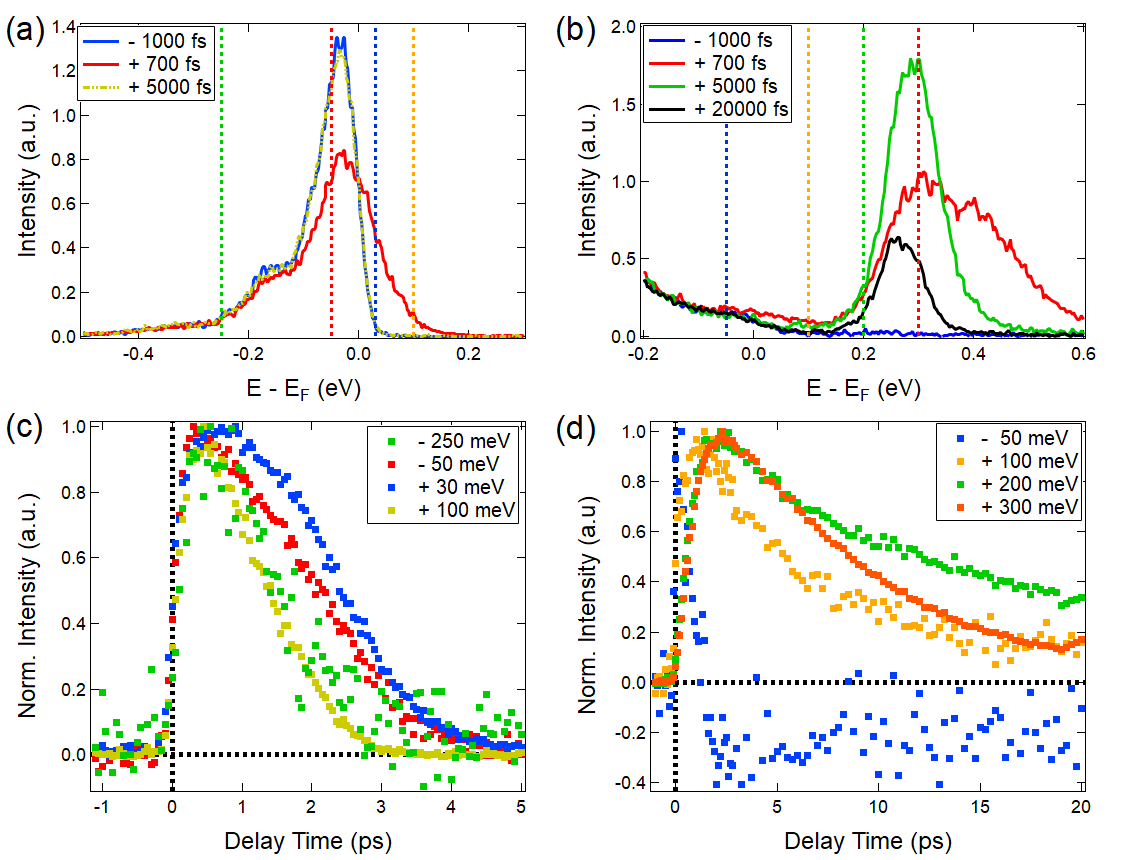}
\caption{Analysis on trARPES results for both \MBT{} and \MBSTpsev{} at 5~K. (a),(b) EDC's at $\Gamma$ at various delay times in the undoped and doped cases respectively. The curves were integrated over a 100~fs delay time range in order to reduce the noise. (c),(d) Normalized, k-integrated intensities as a function of delay time for the undoped and doped compounds, respectively. For clarity, the energies at which the decay curves were extracted are marked by their corresponding colors in (a) and (b).}
\label{Fig3}
\end{figure}

We highlight these differences further in figure~\ref{Fig3}, where we show line profiles on the data in \ref{Fig2}. Figures~\ref{Fig3}a,b show the energy distribution curves (EDC's) taken at the $\Gamma$ point at the delay times presented in figure~\ref{Fig2}(a-f). In \MBT{} at $+700$~fs (figure~\ref{Fig3}a), we observe a decrease in intensity everywhere below E$_F$, and the formation of the high energy tail above E$_F$. After a few picoseconds, the EDC is nearly identical to that before the pump, showing that the system is fully decayed. In the doped sample (figure~\ref{Fig3}b), the majority of changes occur above E$_F$, although there is a small decrease of intensity just below E$_F$. At a $+~700$~fs delay time, there are three prominent features above E$_F$: a tail extending within 0.1~eV of E$_F$, a large peak centered near 0.3~eV, and a broad shoulder beginning near 0.4~eV. We assign these features as the unoccupied part of the lower TSS, a combination of the bulk conduction band and upper TSS, and the equivalent of the Rashba-like state that is in MBT, although we are unable to resolve a Rashba-like dispersion. Importantly, and just as in the case of figure~\ref{Fig1}c,d, we are unable to clearly resolve a Dirac cone due to energy broadening. However, our data suggests that the TSS is gapped, as in the cases of figure~\ref{Fig1}c and Ref.~\cite{Ma2021}.


We now move to discuss the dynamics of both systems. Figures~\ref{Fig3}c,d show the normalized, k-integrated intensity differences at four energies. For clarity, the energies are marked by their corresponding colors in figure~\ref{Fig3}a,b. In MBT(Figure~\ref{Fig3}c), the decay curves correspond to the Dirac point (although the energy range was expanded to include the top of the valence band and bottom of the conduction band to reduce the noise), and inside the Rashba-like state at three different energies in the vicinity of E$_F$. The decay curves for all four energies are similar, differing only by $\sim1$~ps, with the populations reaching a maximum 500~fs after the pump pulse. 

For \MBSTpsev{} (Fig.~\ref{Fig3}d), we clearly observe a strong energy dependence to the decays, in stark contrast with MBT. The representative energies here are at the top of the valence band, inside the bulk gap, at the bottom of the conduction band, and in the middle of the conduction band. The slowest rate is at the bottom of the conduction band, and agrees with the data presented in Fig.~\ref{Fig3}b, where only the bottom of the conduction band is visible. The electron population also reached a maximum 2000~fs after the pump pulse, much later than that in \MBT{}. This agrees with work done on nonmagnetic TI's where longer decays also correspond to longer rise times and are typical when the chemical potential is located inside the bulk gap\cite{Sterzi2017,Hedayat2018,Sumida2017}. Interestingly, we also observe a persistent hole population below E$_F$. These results imply that the electrons and holes are prohibited from recombining due to a lack of decay channels at the Fermi level.

\begin{figure*}[tb]
\includegraphics[width=17.2 cm]{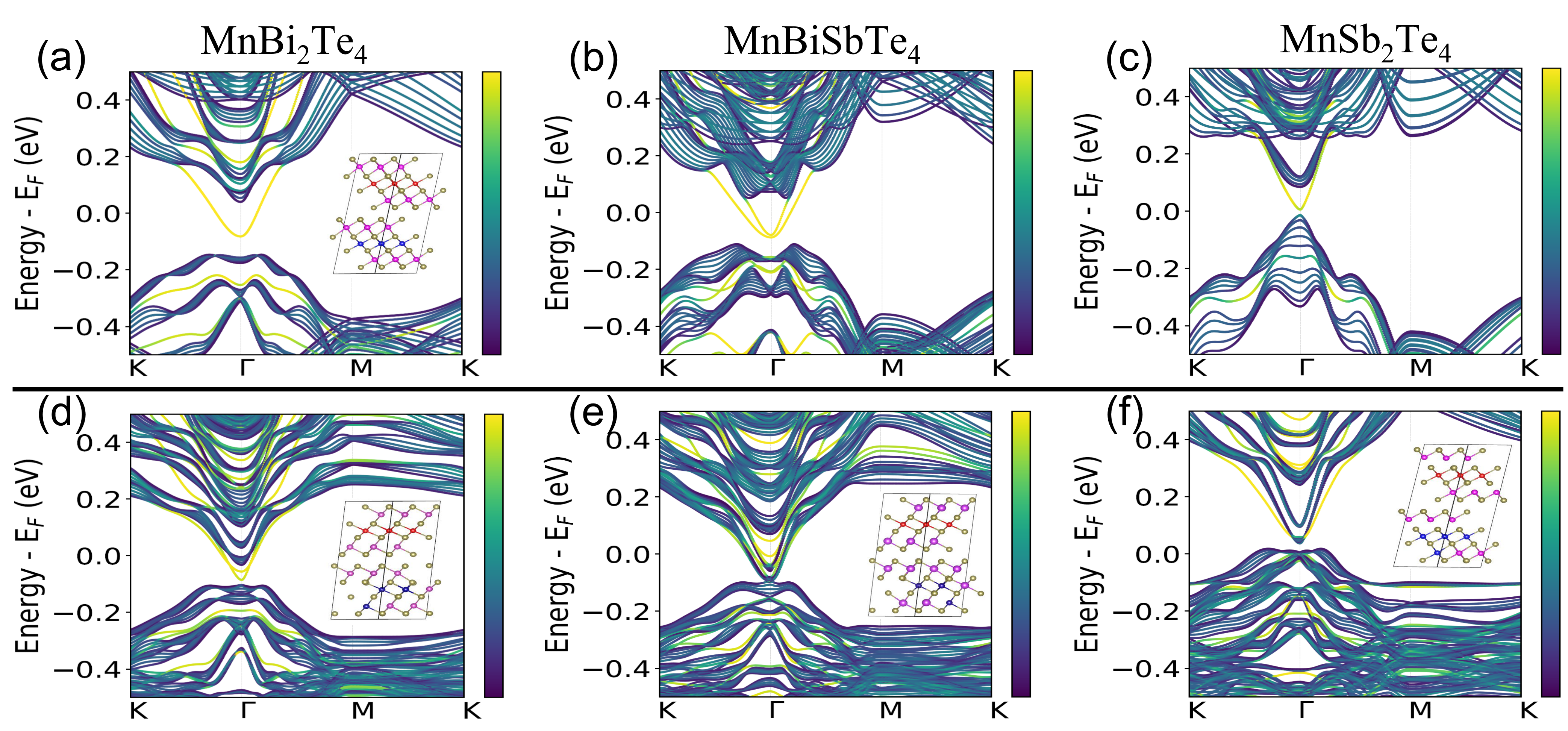}
\caption{DFT band structure calculations. (a-c) Defect-free calculations for 0\%, 50\% and 100\% Sb doping levels respectively. (d-f) Band structures for three different defect configurations of \MBT{}. (d) and (e) are for two different Bi/Mn antisite configurations and (f) includes an Mn$_{Bi}^+$ defect. The color bars represent the projection on the surface, where yellow represents the surface and purple represents the bulk. The insets show the spatial configurations used in the calculations. Yellow and pink are Te and Bi/Sb atoms, and red/blue represent Mn up/down spin species, respectively.}
\label{Fig4}
\end{figure*}

Revealed by our experimental studies, the most important remaining question is why the TSS gap increases with Sb doping concentration. It is likely that this is related to the increase in antisite defect concentration due to the comparable sizes of Sb with Mn, as was previously discussed. Indeed, the broadening of the energy bands, observed in figure~\ref{Fig1}c,d, is evidence of increased defect concentration. This motivates us to perform DFT calculations for the MBT band structure, both as a function of Sb doping, and again in the presence of Mn/Bi antisite defects.


Figure \ref{Fig4} shows the results of such calculations in the presence of A-type AFM order. Figure \ref{Fig4}a-c shows the defect free band structure at 0\%, 50\%, and 100\% Sb doping concentration. All three cases show inverted bulk band gaps and topological surface states, with the gap shrinking with increasing Sb concentration. Aside from the gap sizes, there does not appear to be significant differences between the calculated bands in Figures~\ref{Fig4}a-c. Here, for our defect calculations, we only show our results for MBT for simplicity. However, more results are presented in Ref[supplemental].

Figures~\ref{Fig4}d-f show our calculations for three different antisite defect configurations. Figures~\ref{Fig4}d,e show calculated band structures where an Mn and Bi were swapped at different locations, as shown in their respective insets, and figure~\ref{Fig4}f shows the band structure for a single Mn$_{Bi}^+$ defect. We see that all three calculated defect band structures are different, demonstrating just how sensitive the band structure is to defects. Thus even for samples with similar defect concentrations and types, the band structures may not be the same. Indeed, this was the case in ref\cite{Shikin2021} where samples grown by the same method were shown to have different surface gaps. 

Another important observation is that in all cases, the bulk band gap has decreased, and in the case of figure~\ref{Fig4}e, the bulk band gap has closed. This is an important observation because when the bulk band gap shrinks, the topological surface state becomes more bulk-like. In other words, the penetration depth increases in the presence of defects! Thus, the properties of the topological state are greatly impacted by antisite defects. A final, subtle point is that antisite defects, as shown in figures~\ref{Fig4}d,e, shift the bulk conduction band below the Fermi level, even though the composition is still stoichiometric. Thus antisite defects may not only explain certain properties of the topological surface state, but also some of the intrinsic n-doping. As a final point, we note that our calculations consider specific antisite defects, where in reality, the system will contain a combination of many possible antisite configurations, in addition to domains of perfect crystal.


In conclusion, we have performed a combined experimental and theoretical investigation of the effects of Sb doping and antisite defects. Our photoemission results show that at an Sb concentration of 35\%, the Fermi level is inside the bulk gap. This is  evident in the electronic structures observed with static photoemission, and supported by the long population lifetime in time-resolved photoemission. Examination of the unoccupied band structure in the charge neutral material, as well as the occupied band structure at a lower doping level, show the unanticipated consequence that Sb doping widens the topological gap, complicating the utilization of these materials for device applications. Finally, DFT calculations in the presence of antisite defects support our experimental hypothesis that they can strongly effect the properties of the TSS. Thus defect control is crucial if these materials are to be utilized for QAHE device applications.

Work at Brookhaven National Laboratory was supported by the U.S. Department of Energy, Office of Science, Offce of Basic Energy Sciences, under Contract No. DE-SC0012704. Work at Oak Ridge National Laboratory (ORNL) was supported by the U.S. Department of Energy, Office of Science, Materials Sciences and Engineering Division. H. M. is supported by the Laboratory Directed Research and Development (LDRD) of ORNL, under project No. 10018. The work at Howard University was supported by the US Department of Energy (DOE), Office of Science, Basic Energy Sciences Grant No. DE-SC0022216.

\bibliography{ref}
\end{document}